%% file: main_arxiv.tex
\documentclass[twocolumn,secnumarabic, amssymb, aps, prl, longbibliography]{revtex4-2}
\usepackage{lipsum}
\usepackage{xcolor}
\usepackage{changes}
\usepackage{graphicx}
\usepackage{booktabs}

\usepackage{siunitx}
\usepackage{tikz}
 
\usepackage{ulem}
\usepackage{natbib}
\usepackage{mathtools}
\usepackage[capitalise]{cleveref}
\usepackage{xspace}
\usepackage{braket}
\usepackage{comment}
\usepackage{dcolumn}

\usepackage{makecell}
\usepackage{tabularx}
\newcolumntype{Y}{>{\centering\arraybackslash}X}

\usepackage{multirow}

\newcommand{\fm}[1]{\ifmmode#1\else$#1$\fi}

\newcommand{\Ti}{\fm{^{48}\text{Ti}^{+}}\xspace}
\newcommand{\Ca}{\fm{^{40}\text{Ca}^{+}}\xspace}

\newcommand{\CaS}{\fm{^2\text{S}_{1/2}}\xspace}
\newcommand{\Fa}{\fm{\text{a}^{4}\text{F}}\xspace}

\begin{document}
\bibliographystyle{plainnat}

\title{Land\'e $g$ factor measurement of $^{48}$Ti$^+$ using simultaneous co-magnetometry and quantum logic spectroscopy}

\author{Till Rehmert$^{1,2}$}%
\email{till.rehmert@ptb.de}
\author{Maximilian J. Zawierucha$^{1,2}$}%
\author{Kai Dietze$^{1,2}$}
\author{Piet O. Schmidt$^{1,2}$}%
\author{Fabian Wolf$^1$}%
\email{fabian.wolf@ptb.de}
\affiliation{$^1$Physikalisch-Technische Bundesanstalt, Bundesallee 100, 38116 Braunschweig, Germany\\
$^2$Institut für Quantenoptik, Leibniz Universität Hannover, Welfengarten 1, 30167 Hannover}
\author{Sergey G. Porsev}
\author{Dmytro Filin}
\author{Charles Cheung}
\author{Marianna S. Safronova}
\affiliation{Department of Physics and Astronomy, University of Delaware, Delaware 19716, USA}

\date{\today}
\begin{abstract}
	The use of atomic systems as accurate magnetic field probes requires precise characterization of the particle's magnetic properties. Insufficient knowledge of the spatial and temporal characteristics of the external magnetic field often limits the determination of the corresponding atomic parameters. Here, we present a quantum logic scheme mitigating systematic effects caused by temporal magnetic field fluctuations through simultaneous co-magnetometry. This allows measurement of the ground state $g$ factors of a single \Ti ion with uncertainties at the $10^{-6}$ level. We compare experimentally determined $g$ factors with new theoretical predictions using a combination of configuration interaction (CI) and second-order many-body perturbation theory (MBPT). Theory and experiment agree within the expected level of accuracy. The scheme can be applied to many atomic species, including those that cannot be directly laser cooled.
	
\end{abstract}
\maketitle
Land\'e $g$ factors quantify magnetic moments associated with angular momenta of particles. 
Precise knowledge of these allows the use of atomic systems as magnetic field probes in laboratory~\cite{farooq_absolute_2020,shen_simple_2014,baumgart_ultrasensitive_2016,ruster_entanglement-based_2017} and astrophysical settings~\cite{judge_atomic_2017,kawka_cool_2011}. Currently, the most precise $g$ factor measurements are performed in Penning traps~\cite{werth_zeeman_2018}.
There, $g$ factors can be measured independent of the precise magnetic field strength by relating it to the trapping frequencies. This approach mitigates uncertainties due to magnetic field fluctuations, which would otherwise limit the achievable precision, allowing for determinations of $g$ with an uncertainty of $<10^{-9}$~\cite{brown_precision_1982, sturm_electron_2013}. However, Penning traps operate at field amplitudes of several Tesla to confine the charged particle. Such high magnetic fields prevent the measurement of $g_J$ factors for singly charged ions with weak $LS$-angular-momentum-coupling, since orbital angular momentum ($L$) and electronic spin ($S$) are decoupled in the Paschen-Back regime. Furthermore, the extended averaging times required for these measurements in Penning traps hinder the accurate measurement of short-lived metastable states. Consequently, many atomic states lack precise $g$ factor measurements and rely on theoretical calculations~\cite{li_multiconfiguration_2020,li_mcdhf_2020}.

Here, we present a measurement scheme for accurate determinations of $g$ factors in the weak $B$-field regime. This scheme is independent of the specific atomic species and based on quantum logic in a linear Paul trap. By simultaneously interrogating a co-trapped logic ion with a well-known $g$ factor as a co-magnetometer, our scheme enables precise measurements of the $g$ factor for a wide variety of atomic states. Measurements of $g$ factors relying on well-known reference states in the same species~\cite{hoffman_radio-frequency-spectroscopy_2013, micke_coherent_2020, ma_precision_2024} and in co-trapped species~\cite{rosenband_observation_2007, spies_excited-state_2025, thekkeppatt_measurement_2025} are an established tool to reduce uncertainty due to insufficient knowledge and control over the magnetic field. Going beyond previous implementations, we present a scheme, where the co-magnetometer is not interrogated interleaved but simultaneously with the spectroscopy ion, resulting in suppression of systematic errors due to temporal magnetic field variations. Similar schemes have been demonstrated in Penning traps~\cite{sailer_measurement_2022} as well as neutral atom experiments~\cite{thekkeppatt_measurement_2025} and have been proposed for the characterization and mitigation of systematic errors in optical clocks~\cite{akerman_atomic_2018,wolf_scheme_2024}. We implement this scheme to determine the $g$ factors of all four $\ket{J}$ states of the ground state \Fa of a single \Ti ion using the $^2\text{S}_{1/2}$ ground state in a \Ca ion as a co-magnetometer. 

Titanium is an element of significant astrophysical relevance, as its emission and absorption lines are present in many cosmic spectra such as in stellar spectra~\cite{deb_ti_2009} and quasar absorption spectra~\cite{peroux_most_2006}. This provides valuable information for studies on the variation of fundamental constants~\cite{murphy_laboratory_2013} and stellar composition analysis~\cite{scott_elemental_2015}. Furthermore, titanium is an interesting candidate for testing quantum electrodynamics (QED) within complex atomic systems. By comparing the experimentally obtained $g$ factors to theoretical predictions, this study allows for the exploration of the role of negative energy states~\cite{lindroth_ab_1993} and QED effects in atomic structure calculations of transition metal ions, thus advancing our understanding of fundamental physics in these systems.

\begin{figure}
    \hspace*{0.7cm}
		\sffamily            
		\def\svgwidth{.8\columnwidth}
		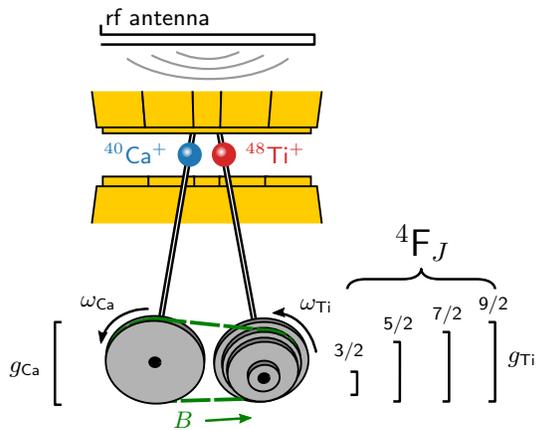
	\caption{Visualization of the measurement principle. A two-ion crystal is trapped in a linear segmented Paul trap. An in-vacuum antenna couples states with frequency splittings in the \SI{}{\kHz} to \SI{}{\MHz} regime. The connection between the magnetic field $B$, $g$ factors and Zeeman level splitting is visualized by a belt drive. In this picture the magnetic field corresponds to the speed of the belt, the radii of the pulleys represent the $g$ factors and their angular velocities $\omega_\text{Ca}$ and $\omega_\text{Ti}$ the Zeeman level splittings for the \CaS and $\ket{\Fa, J}$ manifolds of the logic ion and the spectroscopy ion, respectively. The ratio of the $g$ factors (ratio of the pulleys' radii) can be determined by measuring the Zeeman splittings (angular velocities of the pulleys) irrespective of the magnetic field amplitude $B$.}
	\label{fig:trap_setup}
\end{figure}

The energy shift between adjacent angular momentum projection eigenstates $\ket{m_J}$ and $\ket{m_J+1}$ in a magnetic field $B$ is given by 
\begin{equation}
    \Delta E = g \mu_\text{B} B + g^{(2)}\frac{(\mu_\text{B}B)^2}{m_e c^2}+\mathcal{O}(B^3)
\end{equation}
with the Land\'e $g$ factor, the Bohr magneton $\mu_\text{B}$, the electron mass $m_e$, and the speed of light $c$. $g^{(2)}$ denotes the second order Zeeman coefficient. Since the $g$ factor for the \CaS state in \Ca is precisely known from Penning trap experiments~\cite{tommaseo_mathsfg_scriptscriptstyle_2003}, measurement of the \CaS Zeeman splitting energy $\Delta E_\text{Ca}$ during the experimental sequence determines the magnetic field at the position of the ion (see \cref{fig:trap_setup} for an illustration of the measurement principle). 
Assuming a spatially homogeneous magnetic field, the $g$ factor of \Ti, labelled as $g_\text{Ti}$, can, to first order, be inferred from
 \begin{equation}
    \label{eq:g_transfer}
     g_\text{Ti} = g_\text{Ca} \frac{\Delta E_\text{Ti}}{\Delta E_\text{Ca}},
 \end{equation}
where $g_\text{Ca}$ is the $g$ factor for the co-trapped calcium ground state~\cite{tommaseo_mathsfg_scriptscriptstyle_2003}, and $\Delta E_\text{Ti}$ and $\Delta E_\text{Ca}$ are the Zeeman energy splittings for titanium and calcium, respectively. In \cref{eq:g_transfer}, higher order contributions to $\Delta E$ can be neglected due to sufficiently small $g^{(2)}$ coefficients for the \Ca \CaS and the \Ti \Fa fine structure states. A detailed discussion of the method of calculating $g^{(2)}$ and the resulting corrections to $g_\text{Ti}$ are given in the Supplemental Material~\cite{supplementary}.

\textit{Experimental setup --- }The introduced measurement scheme is realized in a linear segmented Paul trap~\cite{leopold_cryogenic_2019}. A two-ion crystal composed of one \Ca and one \Ti ion is prepared by loading multiple ions of each species and splitting the resulting crystal until the desired configuration is obtained~\cite{zawierucha_deterministic_2024}. A detailed description of the implemented quantum logic methods is given in Ref.~\cite{rehmert_quantum_2025}; in the following the most relevant aspects are summarized. Operations coupling the $\left|\downarrow \right> = \left| S_{1/2}, m_J=-1/2 \right>$ and the $\left|\uparrow \right> = \left| D_{5/2}, m_J=-1/2 \right>$ optical qubit states on the \Ca logic ion are performed with a narrow-linewidth laser at $\SI{729}{\nano\meter}$. A magnetic field of $\SI{0.397}{\milli\tesla}$ defines a quantization axis and results in a \Ca ground state splitting of $\sim\SI{11.135}{\mega\hertz}$. \Ti state preparation and quantum logic readout is facilitated using a far-detuned laser~\cite{chou_preparation_2017, rehmert_quantum_2025} with a wavelength of $\SI{532}{\nano\meter}$ driving Raman transitions. Polarization and orientation of the two Raman beams with respect to the trap axis allow population transfer between adjacent Zeeman states as well as exciting motional modes of the two-ion ensemble.
This way, state information from the spectroscopy ion can be transferred to the shared motional state and read out on the logic ion.
Radio frequency (rf) fields from an in-vacuum antenna can couple Zeeman states on both the calcium and the titanium ion.

\begin{figure}
        \flushleft
        {\normalfont \small \textbf{(a)} }\\\vspace*{-0.1cm}
        \begin{scriptsize}
		\sffamily
		\def\svgwidth{\columnwidth}
		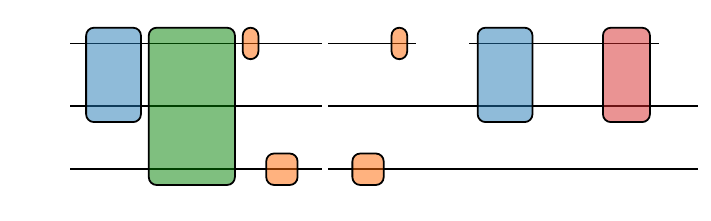\\
        \vspace{10pt}
        {\normalfont \small \textbf{(b)}}
    	\includegraphics[width=\columnwidth]{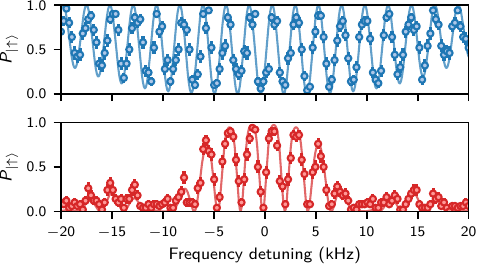}

	\end{scriptsize}
	\caption{Simultaneous rf Ramsey spectroscopy on \Ti and \Ca. \textbf{(a)} Experimental sequence starting with motional ground state cooling (GSC) followed by the preparation of a $\ket{J, m_J=-J}_\text{Ti}$ stretched state in the a$^4$F electronic ground state. Two rf pulses open Ramsey interferometers on the calcium and titanium ion, sequentially. After a common wait time of $\SI{300}{\micro\second}$ two additional pulses close the interferometers. After state detection on the \Ca qubit transition, the \Ti Zeeman state distribution is mapped to \Ca after additional GSC and by a red sideband (RSB) RAP pulse with the Raman laser on \Ti. \textbf{(b)} Ramsey fringes for the \Ca ground state (upper plot) and titanium \Fa ($J=3/2$) state (lower plot). Solid lines are fits to experimental data.} 
	\label{fig:parallel_rf_ramsey_seq}
\end{figure}

\textit{Measurement of \textit{g}-factors --- }The experimental sequence for simultaneously probing Zeeman transitions in the spectroscopy and the logic ion starts with ground state cooling (GSC), realized by a combination of Doppler, electromagnetically-induced-transparency (EIT) cooling and sideband cooling. The preparation of a stretched state ($\left|m_J=\pm J\right>_\mathrm{Ti}$) in one of the four $\Fa_J$ ($J\in\left\{3/2, 5/2, 7/2, 9/2\right\}$) ground states of the \Ti is achieved by population transfer pulses on a common motional red sideband that are rendered irreversible by GSC on \Ca~\cite{rehmert_quantum_2025,schmidt_spectroscopy_2005}. A Ramsey interferometer is opened on both the \CaS state and one of the \Ti \Fa states by applying two rf pulses with individual times of $t_{\pi}^\text{Ca}/2$ ($t_{\pi}^\text{Ti}/2$) with $t_\pi$ being the time required for a rotation of the angular momentum by an angle $\theta = \pi$. After a Ramsey dark time of $t_{R}^\mathrm{Ca}$ ($t_{R}^\mathrm{Ti}$) the two Ramsey interferometers are closed by a $t_\pi/2$ pulse on each species. This Ramsey interferometry sequence converts phase differences between the local rf oscillator and atomic transition frequency to detectable Zeeman population imbalances in \Ca and \Ti Zeeman manifolds. The \Ca state is read out through fluorescence detection (FD) after shelving the Zeeman state information to state $\ket{\uparrow}$ of the optical qubit. The photon scattering during the detection introduces motion to the system but does not change the population distribution in the \Ti Zeeman manifold. Next, the \Ti Zeeman population is read out using quantum logic. This is accomplished by cooling to the motional ground state using the calcium logic ion and mapping the \Ti Zeeman population to a motional mode of the two-ion crystal, which is finally read out at the \Ca ion by a rapid-adiabatic-passage (RAP) pulse on a red sideband of the calcium qubit transition~\cite{gebert_detection_2016}. The full experimental sequence is visualized in \cref{fig:parallel_rf_ramsey_seq}\,(a). In \cref{fig:parallel_rf_ramsey_seq}\,(b), a typical measurement result using $t_{R}^\mathrm{Ti}=\SI{300}{\us}$ and $t_{R}^\mathrm{Ca}=t_{R}^\mathrm{Ti}+t_{\pi}^\text{Ti}$ is shown. A detailed description of the fit function used in \cref{fig:parallel_rf_ramsey_seq}~(b) is provided in the Supplemental Material~\cite{supplementary}.

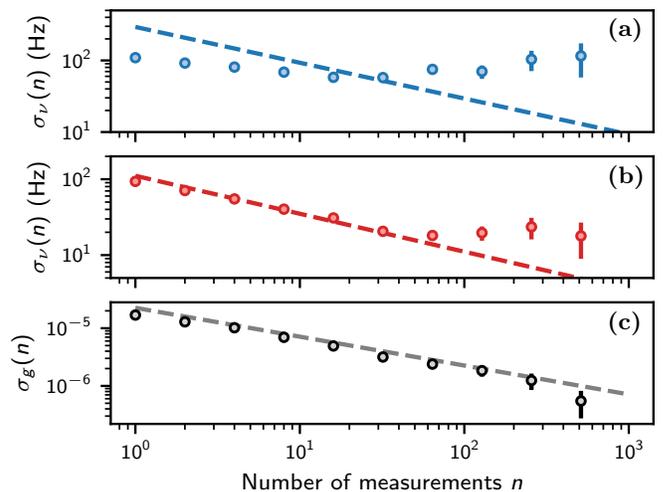
\begin{figure}
\begin{picture}(\columnwidth, 190)
    \put(0, 0){\includegraphics[width=\columnwidth]{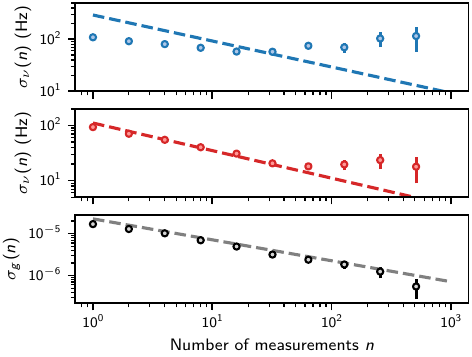}}
    \put(227, 175){\textbf{\small (a)}}
    \put(227, 119){\textbf{\small (b)}}
    \put(227, 64){\textbf{\small (c)}}
\end{picture}	
 \caption{Measurement of the $g$ factor of state $\ket{J=3/2}$ in the \Fa ground state. \textbf{(a), (b)} Allan deviation $\sigma_{\nu}(n)$ of the measured calcium S$_{1/2}$ ground state splitting and the \Ti $\ket{J=3/2, m_J}\rightarrow \ket{J=3/2, m_J-1}$ state splitting, respectively. \textbf{(c)} Allan deviation of the calculated $g$ factor. Dashed lines indicate expected white noise scaling $\sigma(n=1)/\sqrt{n}$.}
	\label{fig:J32_adev}
\end{figure}

To follow magnetic field drifts the applied frequency of the rf pulses was kept on resonance using two individual servo loops~\cite{peik_laser_2006}. For each \Ti ion fine structure state at least 1000 lock iterations, each comprised of 10 measurement repetitions, were recorded. The Zeeman splitting frequency stabilities are analyzed using Allan deviations, depicted in \cref{fig:J32_adev}\,(a) and (b) for the calcium and the titanium ion, respectively. An Allan deviation for quantum projection noise-limited measurements averages down with $\propto\sqrt{1/n}$ corresponding to white frequency noise, where $n$ is the number of measurements. For both, the \Ca and \Ti, white noise behaviour is observed only on short time scales. On longer time scales the frequency variation is dominated by magnetic field fluctuations resulting in a deviation from the $\sqrt{1/n}$-scaling of the Allan deviation. 
However, the recovery of the white frequency noise scaling of the Allan deviation for $g_\mathrm{Ti}$ demonstrates the robustness of the scheme against such fluctuations (see \cref{fig:J32_adev}\,(c)). 
This enables a determination of the $g$ factor with a statistical uncertainty on the $10^{-6}$ level. The final statistical uncertainties are limited by the measurement time (for details see the Supplemental Material~\cite{supplementary}).
Systematic effects, such as the electric quadrupole shift and the non-linear Zeeman effect are estimated to be well below the statistical uncertainties. However, trap-drive induced ac-Zeeman shifts result in a significant shift of the measured $g$ factors. The total systematic and statistical uncertainties are given in \cref{tab:ti-g-factors}. The final accuracy of the measured $g$ factors is limited by the uncertainty of the static magnetic field gradient resulting in different magnetic field strengths for the two-ion positions and the uncertainty in the determination of the oscillating trap magnetic field that causes the ac-Zeeman shift. A detailed discussion of systematic shifts can be found in the Supplemental Material~\cite{supplementary}. The measured $g$ factors for the fine structure of the \Fa ground states are summarized in \cref{tab:ti-g-factors}.

\begin{figure}
        \includegraphics[width=\columnwidth]{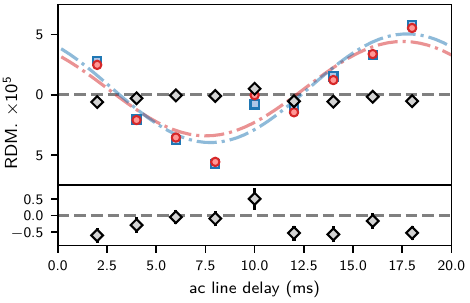}
	\caption{Measurement of the $\ket{J=3/2}_\mathrm{Ti}$ state's $g$ factor in dependence of an additional delay between the ac line trigger and the start of the experimental sequence. Blue squares: relative deviation from the mean (RDM) of all frequency data of the measured \Ca S$_{1/2}$ splitting. Red circles: relative deviation from the mean of all frequency data of the measured Zeeman splitting in the \Ti $J=3/2$ manifold. For the frequency data, errorbars indicate standard error of the mean. Black diamonds: extracted $g$ factor; errors indicate statistical uncertainty propagated from the uncertainty of the frequency measurements. Sine functions were fitted to the frequency data to guide the eye. Lower plot: deviation of the extracted $g$ factor from the mean value (dashed line).}
	\label{fig:J32_trigger_delay}
\end{figure}

A large fraction of the magnetic field noise in this experiment originates from the ac line and is therefore most prevalent at multiples of \SI{50}{\Hz}. Its influence was analyzed by measuring the $g$ factor of the $\ket{J=3/2}_\mathrm{Ti}$ state for different time delays between the start of the experimental sequence and the ac line~\cite{schmidt-kaler_coherence_2003}. As shown in \cref{fig:J32_trigger_delay}, both frequency measurements vary according to the changing magnetic field, while the variation of the extracted $g$ factor is strongly suppressed. This showcases the success of the co-magnetometry Ramsey scheme.

\begin{table*}
    \caption[\Ti g-factors]{Theoretical and experimental values of the \Ti $g$ factors. $g_{LS}$ denotes $g$ factor values calculated taking only $LS$-coupling into account. Statistical uncertainties were derived from Allan deviations of the calculated data for the $g$ factors. Errors given in experiment column are the statistical and systematic uncertainties.}
    \begin{ruledtabular}
    \begin{tabular}{cdddddd}
        State & \multicolumn{1}{c}{$g_{LS}$} & \multicolumn{1}{c}{theory~\citep{li_multiconfiguration_2020,li_mcdhf_2020}} & \multicolumn{1}{c}{theory [this work]} & \multicolumn{1}{c}{experiment [this work]}\\ 
		\colrule
		$\Fa_{3/2}$&  0.39990 & 0.39853 & 0.3986 & 0.3984617(5)_\text{stat}(244)_\text{sys}\\
		$\Fa_{5/2}$&  1.02858  & 1.02840 & 1.0284 & 1.028318(2)_\text{stat}(51)_\text{sys}\\
		$\Fa_{7/2}$&  1.23813  & 1.23839 & 1.2384 & 1.238325(2)_\text{stat}(53)_\text{sys}\\
		  $\Fa_{9/2}$&  1.33339  & 1.33388 & 1.3339 & 1.333823(4)_\text{stat}(52)_\text{sys}\\

    \end{tabular}
	\label{tab:ti-g-factors}
    \end{ruledtabular}
\end{table*}

\textit{Theory --- }In addition to the experimental determination of the \Fa $g$ factors, we have performed theoretical calculations using an approach combining the configuration interaction~(CI) with the second-order many-body perturbation (MBPT) method~\cite{DzuFlaKoz96,2025pCI} (see Supplemental Material~\cite{supplementary} for more information). 
The CI+MBPT method allows us to account for explicitly not only valence-valence correlations (as a pure CI or the many-configurational Dirac-Fock method used in Ref.~\cite{li_multiconfiguration_2020} does) but also core-valence correlations. Our analysis demonstrates a low sensitivity of the $g$ factors presented in \cref{tab:ti-g-factors} to the core-valence correlations. The $g$ factors obtained within the framework of CI+MBPT are only slightly (by $\sim(3-5)\times10^{-5}$) larger than the pure CI values.
Our work demonstrated, that these core-valence correlations already contribute at that level of accuracy.
This implies that the additional digits reported in Ref.~\cite{li_multiconfiguration_2020} were not intended to be taken as significant, as no theoretical uncertainty was provided.

Our results show only a small discrepancy with the experimental data. This finding suggests that higher-order quantum electrodynamical (QED) contributions and the effects of negative-energy states, neither of which are included, play a crucial role at the level of the achieved experimental precision. In few-electron systems such as highly charged ions it has been shown that the inclusion of these contributions lead to a better agreement of the theory with the experimental data~\cite{spies_excited-state_2025}.
The theoretical and experimental values for the $g$ factors from this work as well as the theoretical values from Ref.~\cite{li_multiconfiguration_2020,li_mcdhf_2020} are listed in \cref{tab:ti-g-factors}.

\textit{Conclusion --- }We have measured the Landé $g$ factors of a single ion with a relative precision $<10^{-5}$ by using a simultaneous Ramsey interrogation on a Zeeman transition of a co-trapped logic ion in a Paul trap.
The low uncertainty is enabled by a strong suppression of magnetic field noise-induced systematic errors using this co-magnetometry scheme. The presented method is transferable to other atomic species and allows to put strong bounds on QED and negative-energy states contributions for systems where theory and experiment achieve similar accuracy. In addition, the precise knowledge of magnetic properties in atomic species is relevant for the determination of systematic effects in many applications such as optical atomic clocks or in high precision spectroscopy in general. Further, we provide new theoretical values for the $g$ factors and see a good agreement with other theoretical predictions~\cite{li_multiconfiguration_2020}.
This measurement constitutes the first $g$ factor measurement of the \Ti ground state.

\section*{Acknowledgments}
We thank Martin Steinel for helpful comments on the manuscript and Jonathan Morgner for discussions on Penning trap $g$ factor measurements. Furthermore, we thank the aluminium optical clock team for providing stable light at $\SI{729}{\nano\meter}$.
This research was funded by the Deutsche Forschungsgemeinschaft (DFG, German Research Foundation) – Project-ID 274200144 – SFB 1227 (DQ-mat) B05 with partial support from Germany's Excellence Strategy EXC-2123 QuantumFrontiers 390837967. The project has received funding from the European Research Council (ERC) under the European Union’s Horizon 2020 research and innovation programme (grant agreement No 101019987).

The calculations in this work were done through the use of Information Technologies resources at the University of Delaware, specifically the high-performance Caviness and DARWIN computer clusters. 
The theoretical work has been supported in part by the US NSF Grant  No. PHY-2309254, OAC-2209639, US Office of Naval Research Grant N000142512105  and by the European Research Council (ERC) under the Horizon 2020 Research and Innovation Program of the European Union (Grant Agreement No. 856415).

\end{document}


\bibliographystyle{plainnat}

\title{Supplemental Material: Land\'e $g$ factor measurement of $^{48}$Ti$^+$ using simultaneous co-magnetometry and quantum logic spectroscopy}

\author{Till Rehmert$^{1,2}$}%
\email{till.rehmert@ptb.de}
\author{Maximilian J. Zawierucha$^{1,2}$}%
\author{Kai Dietze$^{1,2}$}
\author{Piet O. Schmidt$^{1,2}$}%
\author{Fabian Wolf$^1$}%
\email{fabian.wolf@ptb.de}
\affiliation{$^1$Physikalisch-Technische Bundesanstalt, Bundesallee 100, 38116 Braunschweig, Germany\\
$^2$Institut für Quantenoptik, Leibniz Universität Hannover, Welfengarten 1, 30167 Hannover}
\author{Sergey G. Porsev}
\author{Dmytro Filin}
\author{Charles Cheung}
\author{Marianna S. Safronova}
\affiliation{Department of Physics and Astronomy, University of Delaware, Delaware 19716, USA}

\date{\today}
\begin{abstract}
	In this document, we provide supplementary information on the measurements and calculations presented in the main manuscript.
	
\end{abstract}
\maketitle
\tableofcontents
\section{Ramsey spectroscopy in Zeeman manifolds}
\label{app:ramsey_qudit}

In the following, the interference pattern measured for Ramsey interferometer in the calcium and titanium Zeeman manifolds will be described. The population in the calcium ground state $\left|\downarrow \right> = \left| S_{1/2}, m_J=-1/2 \right>$ after the second Ramsey pulse on the $\ket{m_J=-1/2}\rightarrow\ket{m_J=+1/2}$ Zeeman transition can be described by
\begin{align}
\begin{split}
\label{eq:general_ramsey}
    P_{\ket{m_J=-1/2}}
    &= \left|C_{-1/2}\right|^2\\
    &= 1-\left|C_{+1/2}\right|^2\\
    &=|a_R|^2,
\end{split}
\end{align}
where we define
\begin{align}
\begin{split}
\label{eq:ramsey}
    |a_R|^2 &= 1-4\frac{\Omega^2}{\bar{\Omega^2}}\sin^2\frac{\bar{\Omega}\tau}{2}\biggl(\cos\frac{\bar{\Omega}\tau}{2}\cos\frac{\Delta t_R+\phi}{2}\\
    &\qquad-\frac{\Delta}{\bar{\Omega}}\sin\frac{\bar{\Omega}\tau}{2}\sin\frac{\Delta t_R+\phi}{2}\biggr)^2.
\end{split}
\end{align}
as the general solution for a Ramsey experiment~\cite{ramseyMolecular1950}.
In \cref{eq:general_ramsey} and \cref{eq:ramsey}, $\tau=t_\pi/2$ denotes time for a $\pi/2$-pulse on the Zeeman transition and $t_R$ the Ramsey dark time. An additional phase shift between the first and the second $\pi/2$-pulse is denoted by $\phi$. $\bar{\Omega}=\sqrt{\Omega^2+\Delta^2}$ is the effective Rabi-frequency for a detuning $\Delta$.

In a two-level system, oscillating magnetic fields can be used to rotate the spin state of an atom.
Such rotations can be expressed in terms of
\begin{equation}\label{eq:H_rf}
    \bm{\hat{H}} = \bm{\gamma}\cdot\bm{\hat{S}}=\gamma_x\hat{\sigma}_x + \gamma_y\hat{\sigma}_y + \gamma_z\hat{\sigma}_z,
\end{equation}
with magnetic vector $\bm{\gamma}$. The spin Pauli matrices $\hat{\sigma}_i$ are the generators of a $\mathrm{SU}(2)$ algebra.
This allows us to describe the dynamics of a $N(=2(J+1))$-level system with total angular momentum $J$ in terms of the $N$-dimensional representation of the unitary group $D^{(J)}$~\cite{hioe_n-level_1987,hamermesh_group_1962} by
\begin{equation}\label{eq:c_m}
\begin{split}
    C_m^{(J)}(t) = \sum_{m'=-J}^{J}D_{mm'}^{(J)}[a, b]C_{m'}(0),\\
    m=-J, -J+1, ...,J,
\end{split}
\end{equation}
with probability amplitude $C_m^{(J)}$ for a certain state $m$.
In \cref{eq:c_m}, $a$ and $b$ describe the two-state time evolution and depend on the particular choice of detuning and Rabi-frequency pulses~\cite{hioe_n-level_1987}.
Starting from an initially prepared streched state ($m = -J$) the occupation probability in state $m$ after the complete evolution of the system becomes
\begin{equation}\label{eq:c_m_short}
    \left|C_{m}^{(J)}\right|^2=\binom{2J}{J-m}\left|a\right|^{2(J-m)}\left|b\right|^{2(J+m)}.
\end{equation}
This model assumes that all levels are equally spaced~\cite{rehmert_quantum_2025} and that the population is only transferred between adjacent ones.
For state $m=-J$, \cref{eq:c_m_short} then simplifies to
\begin{align}
\begin{split}
\left|C_{m=-J}^{(J)}\right|^2 &= \left|a\right|^{4J}.
\end{split}
\end{align}

In a Ramsey experiment implemented through oscillating magnetic fields, the spin dynamics can be fully described by operations satisfying \cref{eq:H_rf}.
This enables us to express the population in the initially prepared state $m=-J$ using \cref{eq:ramsey}.
\begin{align}
\begin{split}
\left|C_{m=-J}^{(J)}\right|^2 &= \left|a_R\right|^{4J}.
\end{split}
\end{align}

\section{Details on Theory}
\label{method}
We consider Ti$^+$ as an ion with three valence electrons above the closed core $[1s^2,\,.\,.\,.\,,3p^6]$. We start from the solution of the Dirac-Hartree-Fock (DHF) equations in the $V^{N-3}$ approximation, where $N$ is the total number of electrons. The initial self-consistency procedure was carried out for the core electrons, and then the $3d,4s$, and $4p$ orbitals were constructed in the frozen-core potential.

The remaining virtual orbitals were formed using a recurrent procedure described in Refs.~\cite{KozPorFla96,KozPorSaf15}, where the large component
of the radial Dirac bispinor, $f_{n'l'j'}$, was obtained from a previously constructed function $f_{nlj}$ by
multiplying it by $r^{l' - l}\, \sin(kr)$, where $l'$ and $l$ are the orbital quantum numbers of the new and old orbitals ($l' \geq l$) and the coefficient $k$ is determined by the properties of the radial grid. The small component $g_{n'l'j'}$ was found from the kinetic balance condition.

The newly constructed functions were then orthonormalized to the functions of the same symmetry.
In total, the basis set included six partial waves ($l_{\rm max} = 5$) and orbitals with the principal quantum number $n$ up to 30.
We included the Breit interaction on the same footing as the Coulomb interaction at the stage of constructing the basis set.

We use an approach that combines the configuration interaction (CI), which considers the interaction between valence electrons, and a method that accounts for core-valence correlations~\cite{DzuFlaKoz96,SafKozJoh09,2025pCI}. The wave functions and energy levels of the valence electrons were found by solving the multiparticle relativistic equation~\cite{DzuFlaKoz96},
\begin{equation}
(H_{\rm FC} + \Sigma(E_n))\, \Phi_n = E_n \Phi_n,
\label{Heff}
\end{equation}
where $H_{\rm FC}$ is the Hamiltonian in the frozen-core approximation and
the energy-dependent operator $\Sigma$ accounts for virtual excitations of the core electrons.
We constructed it using second-order many-body perturbation theory (MBPT) over the residual Coulomb interaction~\cite{DzuFlaKoz96}.
In the following, we refer to this approach as the CI+MBPT method.

The set of Ti$^+$ configurations for even-parity states was constructed by allowing single, double, and triple excitations from the main configuration
of the ground state $3d^2 4s$ to the 4-14$s$, 4-14$p$, 4-14$d$, 4-14$f$, and 5-14$g$ shells (we designate it as $[14spdfg]$).
\subsection{Energy levels}
\label{energies}
The energies of the lowest-lying states of Ti$^+$ are listed in~\tref{Energ}. The energies of the excited states (in cm$^{-1}$) are the offsets from the ground state.
\begin{table}[tp]
\caption{The excitation energies (in cm$^{-1}$) calculated in the pure CI and CI+MBPT approximations are presented.
The experimental values from the NIST database~\cite{RalKraRea11} are given in the last column.}
\label{Energ}%
\begin{ruledtabular}
\begin{tabular}{ccccc}
Config. &\multicolumn{1}{c}{Term} & \multicolumn{1}{c}{CI} & \multicolumn{1}{c}{CI+MBPT} & \multicolumn{1}{c}{Ref.~\cite{RalKraRea11}} \\
\hline      \\[-0.6pc]
$3d^2 4s$ & $a\,^4\!F_{3/2}$ &     0   &    0   &    0   \\[0.1pc]
          & $a\,^4\!F_{5/2}$ &     61  &    96  &    94  \\[0.1pc]
          & $a\,^4\!F_{7/2}$ &    145  &   230  &   226  \\[0.1pc]
          & $a\,^4\!F_{9/2}$ &    253  &   404  &   393  \\[0.1pc]
$3d^3$    & $b\,^4\!F_{3/2}$ &  13884  &   896  &   908  \\[0.1pc]
          & $b\,^4\!F_{5/2}$ &  13927  &   974  &   984  \\[0.1pc]
          & $b\,^4\!F_{7/2}$ &  13985  &  1082  &  1087  \\[0.1pc]
          & $b\,^4\!F_{9/2}$ &  14050  &  1217  &  1216
\end{tabular}
\end{ruledtabular}
\end{table}

In the third and fourth columns, we present the pure CI and CI+MBPT values. The experimental values from the NIST database~\cite{RalKraRea11} are given in the last column. As seen, the energies of the $3d^3\,\,b\,^4\!F_J$
manifold obtained in the framework of the pure CI differ essentially (by 13000 cm$^{-1}$) from the experimental values. Good agreement of these energies with the experiment occurs at the CI+MBPT stage when we include core-valence correlations.
This means that the sensitivity of the $b\,^4\!F_J$ manifold to the core-valence correlations is noticeably higher than the sensitivity of the $a\,^4\!F_J$ states.
\subsection{$g$ factor}
\label{gfactor}
The theoretical values for the $g$-factors are listed in \cref{tab:gfactor}.
To calculate them, we use the operator for the interaction with the external homogeneous magnetic field $\bm{B}$, which is assumed to be directed along the $z$-axis:
\begin{equation}
\label{Vm}
V_m = {\mu}_z  {B}_z = \frac{ec}{2} \,  \sum_{i} (\bm{r}_i \times \bm{\alpha}_i)_z B_z \, ,
\end{equation}
where $\bm{\alpha}_i$ is the vector of the Dirac matrices for the $i$'s electron of the ion, $e$ is the elementary charge, $c$ is the speed of light, and the summation goes over all electrons of the ion.
\begin{table}
\caption{\label{tab:gfactor}  Contributions to the $g$-factors of the low-lying states. The ``CI+MBPT'' results are given in the second column. QED$_\textrm{m}$ denotes contributions 
from the QED corrections to the atomic magnetic moment.}
\begin{ruledtabular}
\begin{tabular}{ccccc}
\multicolumn{1}{c}{State}& \multicolumn{1}{c}{CI+MBPT} & \multicolumn{1}{c}{QED$_{\rm m}$}
                         &  \multicolumn{1}{c}{Final}  & \multicolumn{1}{c}{Exp.}  \\
\hline \\[-0.6pc]
     $a\,^4\!F_{3/2}$    &  0.40001 &  -0.00139 & 0.3986 & 0.3984622(9)  \\[0.1pc]
     $a\,^4\!F_{5/2}$    &  1.02836 &   0.00007 & 1.0284 & 1.028318(2)  \\[0.1pc]
     $a\,^4\!F_{7/2}$    &  1.23786 &   0.00055 & 1.2384 & 1.238326(2)  \\[0.1pc]
     $a\,^4\!F_{9/2}$    &  1.33313 &   0.00077 & 1.3339 & 1.333822(3)  \\[0.3pc]
     $b\,^4\!F_{3/2}$    &  0.40005 &  -0.00139 & 0.3987 &              \\[0.1pc]
     $b\,^4\!F_{5/2}$    &  1.02845 &   0.00007 & 1.0285 &              \\[0.1pc]
     $b\,^4\!F_{7/2}$    &  1.23789 &   0.00055 & 1.2384 &              \\[0.1pc]
     $b\,^4\!F_{9/2}$    &  1.33307 &   0.00077 & 1.3338 & 1.333769(4)  
\end{tabular}
\end{ruledtabular}
\end{table}

A notable contribution to the $g$-factors comes from the QED corrections to the dipole magnetic moment operator and can be estimated as an expectation value of the operator
\begin{equation}
 \Delta \mu_z \equiv \mu_{\rm B} \, \frac{\alpha}{2\pi} \, \sum_i \beta_i \Sigma_{z,i},
\label{del_mu}
\end{equation}
where $\alpha$ is the fine-structure constant, 
$\mu_{\rm B} = e \hbar/2 m_e$ is the Bohr magneton, $\hbar$ is the Planck constant, $m_e$ is the electron rest mass,
$\beta$ is the Dirac matrix,
$\Sigma_z =
\begin{pmatrix}
\sigma_z  & 0 \\
    0     & \sigma_z
\end{pmatrix}$,
and $\sigma_z$ is the Pauli matrix. The respective contributions to the $g$-factors are given in the column labeled ``QED$_{\rm m}$'' in Table~\ref{tab:gfactor}.

By comparing the theoretical and experimental results, we see good agreement. The operator $V_m$ given by \eref{Vm} mixes the large and small components of the wave functions.  In this case, the contribution of negative-energy states can be non-negligible and should be accounted for. The QED corrections are taken into account in the leading approximation. Smaller QED effects should also be included for better theoretical accuracy. We attribute the small remaining discrepancy between theory and experiment to these two factors.

\section{Statistical error determination}
\begin{figure}
    \includegraphics[width=\columnwidth]{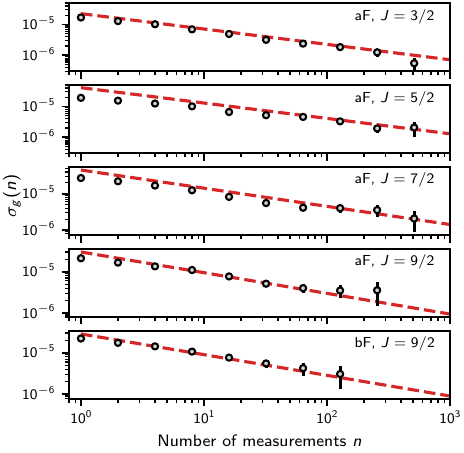}
    \caption{Allan deviations of the derived $g$ factors of the \Ti $\ket{J}$-states for $\Fa_J, J \in \left\{3/2, 5/2, 7/2, 9/2\right\}$ and state $\Fb_{9/2}$. Dashed lines indicate expected white noise behaviour scaling $\propto \sqrt{1/n}$.}
	\label{fig:all_g_adev}
\end{figure}

For each measurement of the Zeeman splitting frequencies in calcium and titanium, the titanium's $g$ factor was calculated from
 \begin{equation}
    \label{eq:g_transfer}
     g_\text{Ti} = g_\text{Ca} \frac{\Delta E_\text{Ti}}{\Delta E_\text{Ca}}.
 \end{equation}
The stabilities of the derived $g$ factors were analyzed using Allan deviations, depicted in \cref{fig:all_g_adev}. For quantum projection noise-limited measurements exhibiting white frequency noise, Allan deviations average down with $\propto\sqrt{1/n}$, where $n$ is the number of measurements. For a single measurement ($n = 1$) the Allan deviation is equivalent to the standard deviation of the collected data defined by
\begin{equation}
    \sigma_{g_\text{Ti}} = g_\text{Ti}\sqrt{\left(\frac{\sigma_{\Delta E_\text{Ca}}}{\Delta E_\text{Ca}}\right)^2 + \left(\frac{\sigma_{\Delta E_\text{Ti}}}{\Delta E_\text{Ti}}\right)^2},
\end{equation}
where $\sigma_{\Delta E_\text{Ca}}$ and $\sigma_{\Delta E_\text{Ti}}$ are the standard deviations of the corresponding measured variables (see \cref{eq:g_transfer}). From \cref{fig:all_g_adev} it can be seen that the Allan deviations follow the expected scaling $\sigma_{g_\text{Ti}}/\sqrt{n}$ (red dashed lines) with the number of measurements. The final statistical uncertainties considered for the titaniums $g$ factors is determined using the Allan deviations calculated at $n = n_\text{max}/2$ measurements, where $n_\text{max}$ represents the total number of measurements for each $J$ state.

\section{Systematic error determination}
\label{app:systematics}

Simultaneous probing of Zeeman transitions in \Ti, $\Delta E_\text{Ti}$, and \Ca, $\Delta E_\text{Ca}$, alleviates temporal magnetic field fluctuations and simplifies the calculation of the $g_\text{Ti}$ factor of the spectroscopy ion according to \cref{eq:g_transfer}.
However, a residual magnetic field gradient along the ion crystal's axis as well as higher order contributions of the Zeeman effect results in additional shifts of the energy levels and thus the measured $g$ factors. These effects are discussed below. The uncertainty associated with $g_\text{Ca}$ from previous measurements~\cite{tommaseo_mathsfg_scriptscriptstyle_2003} is on the order of $4\times 10^{-8}$ and can therefore be neglected.

\subsection{Magnetic field gradient-induced errors}

The two ions separated by a distance $d$ experience a magnetic field difference $\Delta B$. We assume a constant linear gradient. $\Delta B =0.665(1.388)\,\unit{\nano\tesla}$ was derived through measurements of $\Delta E_\mathrm{Ca}$ at different positions in the two-ion crystal.
At the position of the titanium ion the magnetic field is given by $B_\mathrm{Ti} = B_\mathrm{Ca}\pm\Delta B = \Delta E_\mathrm{Ca}/(g_\mathrm{Ca}\mu_\mathrm{B})\pm\Delta B$, where $B_\mathrm{Ca}$ is the magnetic field at the position of the calcium ion. The sign depends on the ion ordering.
The corrected $g$ factor for \Ti can be inferred from $\Delta E_\mathrm{Ti}$ and $\Delta E_\mathrm{Ca}$  by
\begin{align}
    g_\mathrm{Ti}' &= \frac{\Delta E_\mathrm{Ti}}{\mu_\mathrm{B}B_\mathrm{Ti}} \\
    &= g_\mathrm{Ca} \frac{\Delta E_\mathrm{Ti}}{(\Delta E_\mathrm{Ca}\pm g_\mathrm{Ca}\mu_\mathrm{B}\Delta B)}.
\end{align}
For small gradients we can approximate
\begin{align}
    \label{eq:b_gradient_approx}
    g_\mathrm{Ti}' \approx g_0\left(1 \mp \underbrace{\frac{\Delta B}{B_\mathrm{Ca}}}_{\Delta g_\text{grad}/g_0} + \mathcal{O}\left(\frac{\Delta B ^2}{B_\mathrm{Ca}^2}\right)\right),
\end{align}
where we defined $g_0=g_\mathrm{Ca}\frac{\Delta E_\mathrm{Ti}}{\Delta E_\mathrm{Ca}}$, the $g$ factor extracted under the assumption of a spatially uniform magnetic field, and the sign is defined by the order of the ions with respect to the gradient.
The relative shift $\Delta g_\text{grad}/g_0$ resulting from the gradient of the magnetic field is therefore $\Delta B/B_\mathrm{Ca} = \num{1.674e-6}$. An uncertainty of the gradient ($\sigma_{\Delta B}$) arises from the uncertainty of $\Delta E_\mathrm{Ca}$ leading to a systematic uncertainty $\sigma_{\Delta B}/B_\text{Ca}=\num{3.492e-6}$.
Since the uncertainty in the shift is larger than the shift itself, we consider only the uncertainty of the gradient in determining the uncertainty of the $g$ factor.
However, collisions switch the ion positions during the measurements, therefore the first order effect from the approximated gradient (see \cref{eq:b_gradient_approx}) partially averages out and the considered values overestimate the uncertainty.

\begin{figure}
    \includegraphics[width=\columnwidth]{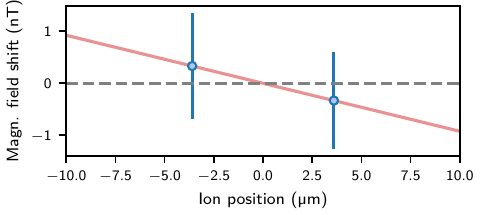}
    \caption{Magnetic field gradient $\Delta B=\SI{0.665(1.388)}{\nano\tesla}$ along the two-ion crystal axis. Blue dots: Absolute magnetic field deviation derived from the measured S$_{1/2}$ ground state splitting of the calcium ion at the two possible ion positions in a two-ion crystal. Errorbars show statistical uncertainty derived from Allan deviations of the calculated field amplitudes. The mean magnetic field was calculated to be $B_\mathrm{Ca}=\SI{0.397}{\milli\tesla}$.}
	\label{fig:b_gradient}
\end{figure}

\begin{table}
\caption{Absolute corrections to $g$ factors due to magnetic field gradient and its uncertainty in brackets.\label{tab:gradientshift}}
\begin{ruledtabular}
\begin{tabular}{cc}
\multicolumn{1}{c}{State}& \multicolumn{1}{c}{$\Delta g_\text{grad}\times 10^{6}$} \\
\hline \\[-0.6pc]
     $a\,^4\!F_{3/2}$    &  0.667(1.392) \\[0.1pc]
     $a\,^4\!F_{5/2}$    &  1.721(3.591) \\[0.1pc]
     $a\,^4\!F_{7/2}$    &  2.073(4.324) \\[0.1pc]
     $a\,^4\!F_{9/2}$    &  2.233(4.658) \\
\end{tabular}
\end{ruledtabular}
\end{table}

\subsection{ac Zeeman shift}
In a static magnetic field, the Zeeman splittings frequencies are given by $\omega = \Delta E/\hbar = g_J \mu_B B/\hbar$. However, an oscillating magnetic field leads to an additional time-averaged energy shift given by~\cite{gan_oscillating_2028}
\begin{equation}
\label{eq:acshift}
    \delta E = \left[\frac{1}{2}\frac{\omega^2}{\omega^2-\Omega^2_\mathrm{rf}}\frac{\braket{B_\perp^2}}{B^2}\right]\omega\hbar,
\end{equation}
where $B_\perp$ is the field amplitude perpendicular to the static field defining the quantization axis and $\braket{\cdot}$ denoting the time average. 
Since trap drive-induced ac magnetic fields affect both, the measurement of $\Delta E_\mathrm{Ca}$ and $\Delta E_\mathrm{Ti}$, the corrected $g$ factor for the different titanium $J$ states can be expressed as
\begin{equation}
    g_\mathrm{Ti}' = g_\mathrm{Ca}\frac{\Delta E_\mathrm{Ti}+\delta E_\mathrm{Ti}}{\Delta E_\mathrm{Ca}+\delta E_\mathrm{Ca}}.
\end{equation}

The Autler-Townes splitting frequency~\cite{autler_stark_1955} in the calcium's S$_{1/2}$ ground state was measured to derive $B_\perp=\SI{3.30(7)}{\micro\tesla}$~\cite{gan_oscillating_2028}. 
The resulting corrections $\Delta g_\mathrm{ac,trap} = g_\mathrm{Ti}-g_\mathrm{Ti}'$ to the measured $g$ factors are summarized in \cref{tab:acshift}. Values in brackets are errors of the derived corrections assuming \SI{1}{\kilo\hertz} accuracy in the Autler-Townes splitting frequency measurement.

Furthermore, we investigated how the rf pulses, which form the Ramsey interferometer on the titanium ion, affect the measurement of $\Delta E_\text{Ca}$.
Here, off-resonant rf pulses induce additional magnetic field shifts in equivalence to laser induced ac light shifts. This leads to an additional phase accumulation $\phi = 2\pi(\omega+\Delta\omega)t_R$ during the dark time $t_R$ of the Ramsey experiment in the calcium ground state. This gives rise to an error in the determination of $\Delta E_\text{Ca}$. We derived the mean magnetic field from the measured Rabi frequencies in Ref.~\cite{rehmert_quantum_2025} weighted by the ratio between the pulse durations $t_\pi^\mathrm{Ti}$ on the titanium ion and the dark time on the calcium ion $t_R^\mathrm{Ca}$:
\begin{equation}
    \braket{B_\perp} = B_\perp \frac{t_\pi^\mathrm{Ti}}{t_R^\mathrm{Ca}}.
\end{equation}
\Cref{eq:acshift} can now be used to calculate the correction $\Delta g_\mathrm{ac,rf}$ to the titanium's $g$ factors arising from the two rf pulses. The obtained values are summarized in \cref{tab:acshift}.
The values for the $g$ factor given in the main text have been corrected considering the ac Zeeman shifts discussed aboved and the uncertainty has been included in the total systematic uncertainty.
\begin{table}
\caption{Absolute corrections to $g$ factors from ac Zeeman shifts. Second column: corrections from trap drive-induced ac Zeeman shift and its uncertainties in brackets. Third column: Corrections from rf pulses forming the Ramsey interferometer on the titanium ion.\label{tab:acshift}}
\begin{ruledtabular}
\begin{tabular}{ccc}
\multicolumn{1}{c}{State}& \multicolumn{1}{c}{$\Delta g_\mathrm{ac, trap}\times 10^5$} & \multicolumn{1}{c}{$\Delta g_\mathrm{ac, rf}\times 10^8$} \\
\hline \\[-0.6pc]
     $a\,^4\!F_{3/2}$    &  2.369(0.104)   & 5.691 \\[0.1pc]
     $a\,^4\!F_{5/2}$    &   4.939(0.216)  & 2.113 \\[0.1pc]
     $a\,^4\!F_{7/2}$    &   5.134(0.225)  & 1.885 \\[0.1pc]
     $a\,^4\!F_{9/2}$    &  5.057(0.222)  & 1.865 \\
\end{tabular}
\end{ruledtabular}
\end{table}

\subsection{Quadrupole shift}
For states with $J>1/2$, the interaction of the electric quadrupole moment with field gradients results in an energetic shift that scales with $m_J^2$. The corresponding Hamiltonian can be written as~\cite{itano_external-field_2000,shaniv_atomic_2016}
\begin{equation}
    \hat H_\mathrm{Q} = \nabla E^{(2)} \hat \Theta^{(2)},
\end{equation}
where $\nabla E^{(2)}$ is the electric field gradient tensor and $\hat \Theta^{(2)}$ is the quadrupole operator.
For the quadrupole field in an ion trap the components $i\in\{0,1,2\}$ of the electric field gradient are well approximated by
$(\nabla E^{(2)})_i = \delta_i\frac{\dif E_z}{\dif z}$ where $z$ indexes the direction of the trap axis.
Therefore, the corresponding energy shift of a state $\ket{\gamma J m_J}$ is given by 
\begin{equation}
\bra{\gamma J m_J}\hat H_\mathrm{Q}\ket{\gamma J m_J} = \frac{\dif E_z}{\dif z} \bra{\gamma J m_J}\hat \Theta_\mathrm{0}\ket{\gamma J m_J} 
\end{equation}
The quadrupole moment $\Theta$ is commonly defined as the expectation value of the operator $\hat \Theta_0$ (the $q=0$ component of the electric quadrupole tensor) in the stretched state $\ket{\gamma J, m_J=J}$, which maximizes the quadrupole shift~\cite{itano_external-field_2000, angel_direct_1967}.
Consequently, the quadrupole shift can be bounded as
\begin{equation}
\Delta E_\mathrm{Q} < \frac{\dif E_z}{\dif z} \Theta.
\end{equation}

The electric field gradient at the position of an ion in a two-ion Coulomb crystal in a linear trap along the axis of dc confinement is given by~\cite{roos_precision_2005}
\begin{equation}
\frac{\dif E_z}{\dif z} = -2\frac{m\omega_z^2}{q},
\end{equation}
where $m$ is the mass of the calcium ion, $\omega_z$ is the axial trapping frequency for a single calcium ion, and $q$ is its charge.
For a trapping frequency of \qty{688}{\kilo\hertz} for a single \Ca this results in $\dif E_z/\dif z = \qty{-1.55e7}{\volt\per\meter\squared} $.
In a Ramsey measurement the quadrupole effect does not result in a shift in first order, due to the population symmetry with respect to the sign of $m_J$. However, an error in Rabi time will result in a frequency shift, since non-perfect $\pi/2$ pulses will result in non-symmetric states. A rough estimate for an upper bound of the error in a $g$ factor measurement for angular momentum state $J$ is given by  
\begin{equation}
    \Delta g_\mathrm{quad}(J) \lesssim\frac{\dif E_z}{\dif z} \frac{\Theta(J)}{\mu_\mathrm{B}B}, 
\end{equation}
which is given in \cref{tab:quadshift}. The quadrupole moments have been calculated theoretically as described below. 

\begin{table}
\caption{\label{tab:quadshift} Upper bounds to corrections to the $g$ factor due to electric quadrupole effects. Electric quadrupole moments $\Theta$ (in $e a^2_{\rm B}$) have been determined theoretically (see also \cref{tab:E2}).}
\begin{ruledtabular}
\begin{tabular}{ccc}
\multicolumn{1}{c}{State}& \multicolumn{1}{c}{$\Theta(J)$} & \multicolumn{1}{c}{$\Delta g_\mathrm{quad}(J)\times 10^7$} \\
\hline \\[-0.6pc]
     $a\,^4\!F_{3/2}$    &  $0.23(3)$   & $\lesssim4.7$ \\[0.1pc]
     $a\,^4\!F_{5/2}$    &  $0.18(3)$   & $\lesssim3.7$\\[0.1pc]
     $a\,^4\!F_{7/2}$    &  $0.17(4)$   & $\lesssim3.5$\\[0.1pc]
     $a\,^4\!F_{9/2}$    &  $0.1$       & $\lesssim2.0$\\
\end{tabular}
\end{ruledtabular}
\end{table}

\subsubsection{Theoretical determination of the quadrupole moments}
The electric quadrupole moment $\Theta$ of an atomic state $| \gamma Jm_J \rangle$ is given by
\begin{eqnarray}
\Theta &=& 2\, \langle \gamma J, m_J=J |Q_0| \gamma J, m_J=J \rangle \nonumber \\
       &=& 2\, \sqrt{\frac{J(2J-1)}{(2J+3)(J+1)(2J+1)}} \langle \gamma J ||Q|| \gamma J \rangle .
\end{eqnarray}
The single-electron electric quadrupole operator is determined as $Q_0 = -e\, r^2 C_{20}({\bf n})$, where ${\bf n} \equiv {\bf r}/r$, and $C_{2q}$ are the normalized spherical 
harmonics ~\cite{VarMosKhe88}. $\gamma$ encapsulates all other quantum numbers. 

We calculated the matrix elements (MEs) of the electric quadrupole operator and quadrupole moments $\Theta$
for the $a\, ^4\!F_J$ and $b\, ^4\!F_J$ states. The values $\Theta$ are presented (in $e a^2_{\rm B}$, where $a_{\rm B}$ is the Bohr radius) in \tref{tab:E2}.
\begin{table}
\caption{\label{tab:E2} Electric quadrupole moments $\Theta$ (in $e a^2_{\rm B}$), obtained in different calculations
(explained in the main text), are presented. Uncertainties are given in parentheses.}
\begin{ruledtabular}
\begin{tabular}{cccccc}
\multicolumn{1}{c}{State}& \multicolumn{1}{c}{pure CI} & \multicolumn{1}{c}{CI+MBPT (I)} & \multicolumn{1}{c}{CI+MBPT(II)}
                         & \multicolumn{1}{c}{Final}\\
\hline \\[-0.6pc]
     $a\,^4\!F_{3/2}$    &   0.35  &   0.23  &   0.20  & $0.23(3)$  \\[0.1pc]
     $a\,^4\!F_{5/2}$    &   0.35  &   0.18  &   0.15  & $0.18(3)$  \\[0.1pc]
     $a\,^4\!F_{7/2}$    &   0.49  &   0.17  &   0.13  & $0.17(4)$  \\[0.1pc]
     $a\,^4\!F_{9/2}$    &   0.74  &   0.12  &   0.06  & $0.1$      \\[0.3pc]
     $b\,^4\!F_{3/2}$    & $-0.69$ & $-0.45$ & $-0.43$ & $-0.45(2)$  \\[0.1pc]
     $b\,^4\!F_{5/2}$    & $-0.67$ & $-0.39$ & $-0.37$ & $-0.39(2)$  \\[0.1pc]
     $b\,^4\!F_{7/2}$    & $-0.93$ & $-0.48$ & $-0.44$ & $-0.48(4)$  \\[0.1pc]
     $b\,^4\!F_{9/2}$    & $-0.68$ & $-0.58$ & $-0.54$ & $-0.58(4)$
\end{tabular}
\end{ruledtabular}
\end{table}

The results obtained within the framework of the pure CI and CI+MBPT methods are given in the columns labeled ``CI'' and ``CI+MBPT (I)'', respectively. Comparing these results, we see that the quadrupole moments
are sensitive to the core-valence correlations. This is especially true for the state $a\,^4\!F_{9/2}$, whose CI+MBPT value differs by a factor of 6 from the CI value. As our analysis shows, in addition, the quadrupole moments of $a\,^4\!F_J$ are sensitive to the mixing of this level to $b\,^4\!F_J$. To reproduce this mixing correctly, one needs to reproduce the energy differences between the $a\,^4\!F_J$ and $b\,^4\!F_J$ states as well as possible.

To test the sensitivity of $\Theta$ to the method of constructing the basis set and size of the CI space, we performed another CI+MBPT calculation, designated in the table as ``CI+MBPT (II)''. For this, we constructed a $B$-spline basis set in the $V^{N-3}$ approximation, increasing the number of partial waves and orbitals in each partial wave. In total, this basis set included six partial waves ($l_{\rm max} = 6$) and orbitals with the principal quantum number $n$ up to 35.
In the CI stage, we allowed single, double, and triple excitations from the main configuration $3d^2 4s$ to $[20spdfg]$.
Thus, the CI space was noticeably more extended in comparison to the CI+MBPT (I) approach.

Based on the difference between the CI+MBPT (I) and CI+MBPT (II) results, we assign uncertainties to the quadrupole moments
of the $a\,^4\!F_J$ states. We note that the value of $\Theta (a\,^4\!F, J=9/2)$, obtained from the first CI+MBPT calculation is twice the value from CI+MBPT (II). For this reason, the uncertainty of this value can be as large as 100\%.

\subsection{Quadratic Zeeman shift}
The cancellation of magnetic field dependence works perfectly only in the linear regime, and higher order effects will lead to a deviation. These deviations arise from the onset of the demixing of LS coupled states into eigenstates of spin $S$ and orbital angular momentum states $L$ when transitioning into the Breit-Rabi regime. Therefore, the magnitude of the effect strongly depends on the fine structure interaction and is therefore absent in the calcium ground state, but needs to be considered for titanium. Considering the next higher order, the energy shift due to the Zeeman effect reads
\begin{equation}
    \label{eq:Zeeman_shift}
    \Delta E_{m_J} = m_J g \mu_B B + g^{(2)}(J,m_J)\frac{(\mu_B B)^2}{m_\text{e} c^2} + \mathcal{O}(B^3).
\end{equation}
Therefore \cref{eq:g_transfer} needs to be modified to
\begin{equation}
     g_\text{Ti} = g_\text{Ca} \frac{\Delta E_\text{Ca}}{\Delta E_\text{Ti}} -\underbrace{\frac{g^{(2)}_\text{Ti}(J,m_J)}{g_\text{Ca}}\frac{\Delta E_\text{Ca}}{m_\text{e} c^2}}_{:= \Delta g_{B^2}},
\end{equation}
where nonlinear effects are only considered on the titanium ion, $g^{(2)}_\text{Ti}(J,m_J)$ is the second order Zeeman coefficient. We use theoretical calculations of $g^{(2)}_\text{Ti}(J,m_J)$ to estimate the error of the $g$ factor determination $\Delta g_{B^2}$ due to the second order Zeeman effect. Details of the calculation are given below. Estimated values for $\Delta g_{B^2}$ are listed in \cref{tab:b2shift}.

\begin{table}
\caption{\label{tab:b2shift} Theoretical values for the second order Zeeman coefficient for \Ti $g^{(2)}_\mathrm{Ti}$ and the resulting correction to the measured $g$-factor. The shift is below the statistical uncertainty of the measurement and therefore not resolved. }
\begin{ruledtabular}
\begin{tabular}{ldd}
State &  \multicolumn{1}{c}{$g^{(2)}_\mathrm{Ti}\times 10^{7}$} &  \multicolumn{1}{c}{$\Delta g_{B^2}\times 10^{7}$} \\
\colrule
$\Fa_{3/2}$ & -2.798(5) & -2.00  \\
$\Fa_{5/2}$ & 0.425(7)& 0.304\\
$\Fa_{7/2}$ & 1.191(6)& 0.853\\
$\Fa_{9/2}$ & 1.252(3)& 0.897

\end{tabular}
\end{ruledtabular}
\end{table}
\subsubsection*{Calculation of the quadratic Zeeman coefficient}
We use the isotope $^{48}$Ti$^+$ with the nuclear spin $I=0$. Since there is no hyperfine interaction, an energy shift due to the quadratic Zeeman effect can be written as
\begin{equation}
\Delta E_{m}^{(2)} = g^{(2)}(J,m_J) \frac{\left(\mu_{\rm B} B\right)^2}{m_ec^2}.
\label{eq:zeeman}
\end{equation}

In Ref.~\cite{PorKozSaf23}, the expression for ac magnetic dipole polarizability was derived, including the summation over intermediate positive- and negative-energy states. Using this approach, we can divide the dimensionless coefficient $g^{(2)}$ into two parts
\begin{equation}
 g^{(2)} = g^{(2)}_{\rm pos} + g^{(2)}_{\rm neg} ,
 \label{eq:g2}
\end{equation}
where the first and second terms stand for the positive and negative-energy states contributions, respectively, and are given by
\begin{eqnarray}
g^{(2)}_{\rm pos} &\equiv& \frac{m_e c^2}{\mu_{\rm B}^2} 
\sideset{}{'}\sum_{\gamma_n J_n}
\frac{|\langle \gamma_n J_n m_J |\mu_z + \Delta \mu_z| \gamma_0 J_0 m_J\rangle|^2}{E_0 - E_n},
\label{eq:g2pos}
\end{eqnarray}

\begin{eqnarray}
g^{(2)}_{\rm neg} &\equiv& \frac{e^2 c^2}{12 \mu_{\rm B}^2} 
\langle \gamma_0 J_0 m_J \,|r^2|\, \gamma_0 J_0 m_J \rangle.
\end{eqnarray}
The prime over $\sum$ in \eref{eq:g2pos} means that the intermediate state 
$|\gamma_n J_n \rangle = |\gamma_0 J_0 \rangle$ should be excluded from the summation.
The operators ${\mu}_z$ and ${\Delta \mu}_z$ are determined by Eqs.~(\ref{Vm}) and (\ref{del_mu}).

The summation in \eref{eq:g2pos} covers all intermediate positive-energy states allowed by the selection rules.
In the following, we assume that $| \gamma_0 J_0 m_J \rangle$ is a
$|a\,^4\!F_J\, m_J \rangle$ state. In the sum over intermediate states, we take into account only the states of the same $a\,^4\!F_J$ manifold. For example, to calculate $g^{(2)}_{\rm pos} (a\,^4\!F_{5/2})$ we take into account only the intermediate states $a\,^4\!F_{3/2}$ and $a\,^4\!F_{7/2}$.

We have verified that the contribution of all other intermediate states is negligible.

In the non-relativistic approximation, the expression for the non-diagonal matrix elements (MEs) of the operator $\mu_z + {\Delta \mu}_z$ is reduced to
\begin{eqnarray}
\label{delta_mu_z}
&& |\langle \gamma_n J_n m_J \,| \mu_z + \Delta \mu_z |\, \gamma_0 J_0 m_J  \rangle|  = \nonumber \\
&& = \mu_{\rm B} \, \left( 1 + \frac{\alpha}{\pi} \right)\,
|\langle \gamma_n J_n m_J \,|S_z|\, \gamma_0 J_0 m_J \rangle|,
\end{eqnarray}
where $\bf S$ is the total spin.

To study the role of relativistic corrections, we calculated the reduced MEs 
$\langle \gamma_n J_n || \mu || \gamma_0 J_0 \rangle$ using the nonrelativistic expression \eref{delta_mu_z} and the relativistic form of the $M1$ operator, given by \eref{Vm}. We compare the results obtained in \tref{tab:M1}.
\begin{table}
\caption{\label{tab:M1} $\langle \gamma_n J_n || \mu || \gamma_0 J_0 \rangle$ (in $\mu_{\rm B}$) calculated in relativistic (``Relativ.'') and nonrelativistic (``Nonrel.'') approximations and the QED correction determined as $\langle \gamma_n J_n || \Delta \mu || \gamma_0 J_0 \rangle$ (in $\mu_{\rm B}$) are presented.
The final results are obtained as relativistic values plus the QED correction.}
\begin{ruledtabular}
\begin{tabular}{ccccc}
                            & \multicolumn{1}{c}{Relativ.} & \multicolumn{1}{c}{Nonrel.} & \multicolumn{1}{c}{QED} & \multicolumn{1}{c}{Final} \\
\hline \\[-0.6pc]
$|\langle a\,^4\!F_{5/2} ||\mu|| a\,^4\!F_{3/2} \rangle|$  & 3.0970 & 3.0984 & 0.0072  & 3.104    \\[0.2pc]
$|\langle a\,^4\!F_{7/2} ||\mu|| a\,^4\!F_{5/2} \rangle|$  & 3.5840 & 3.5857 & 0.0083  & 3.592    \\[0.2pc]
$|\langle a\,^4\!F_{9/2} ||\mu|| a\,^4\!F_{7/2} \rangle|$  & 3.1609 & 3.1623 & 0.0073  & 3.168 
\end{tabular}
\end{ruledtabular}
\end{table}

The table shows that the difference between the relativistic and nonrelativistic results is very small, 0.05\%. The QED correction to the $M1$ operator, calculated using \eref{delta_mu_z}, changes the results by 0.2\%. The final results are obtained as the relativistic values plus the QED correction.

In contrast to the quadrupole moments of $a\,^4\!F_J$, the $M1$ transition amplitudes between the fine-structure states of this manifold are insensitive to the core- and valence-valence correlations. As a result, the quadratic Zeeman shift for these states can be calculated with good accuracy. The calculation was carried out in the framework of the CI+MBPT (I) method for the magnetic quantum number $m_J=3/2$, and the results are presented in \tref{tab:Zeeman}.
\begin{table}[htp]
\caption{\label{tab:Zeeman} Calculated $g^{(2)}(J,m_J=3/2)$ coefficients. Uncertainties are given in parentheses.}
\begin{ruledtabular}
\begin{tabular}{cccc}
\multicolumn{1}{c}{State}& \multicolumn{1}{c}{$g^{(2)}_{\rm pos}\times 10^{7}$} & \multicolumn{1}{c}{$g^{(2)}_{\rm neg}\times 10^{7}$}
                         & \multicolumn{1}{c}{$g^{(2)}\times 10^{7}$} \\
\hline \\[-0.6pc]
     $a\,^4\!F_{3/2}$    & $-2.815(3)$ & $0.018(2)$  & $-2.798(5)$  \\[0.1pc]
     $a\,^4\!F_{5/2}$    & $ 0.407(5)$ & $0.018(2)$  & $ 0.425(7)$  \\[0.1pc]
     $a\,^4\!F_{7/2}$    & $ 1.174(4)$ & $0.018(2)$  & $ 1.191(6)$  \\[0.1pc]
     $a\,^4\!F_{9/2}$    & $ 1.234(1)$ & $0.018(2)$  & $ 1.252(3)$
\end{tabular}
\end{ruledtabular}
\end{table}

We estimate the uncertainty of $g^{(2)}_{\rm pos}$ based on the difference between the relativistic
and nonrelativistic values of the matrix elements of the $M1$ operator.
An additional source of uncertainty is $g^{(2)}_{\rm neg}$. Since $r^2$ is a scalar operator, it is necessary to calculate the contribution of valence and core electrons to $\langle \gamma_0 J_0 m_J \,|r^2|\, \gamma_0 J_0 m_J \rangle$. The core contribution is calculated in the single-electron approximation. We estimate the precision of this approximation to be about 10\%. Since the core contribution gives about 40\% of the total value of $g^{(2)}_{\rm neg}$, we conservatively estimate the uncertainty of $g^{(2)}_{\rm neg}$ at the level of 10\%.
The final uncertainties for $g^{(2)}$ were found as the sum of the uncertainties for $g^{(2)}_{\rm pos}$ and $g^{(2)}_{\rm neg}$.

%% file: fig1.pdf_tex
\begingroup%
  \makeatletter%
  \providecommand\color[2][]{%
    \errmessage{(Inkscape) Color is used for the text in Inkscape, but the package 'color.sty' is not loaded}%
    \renewcommand\color[2][]{}%
  }%
  \providecommand\transparent[1]{%
    \errmessage{(Inkscape) Transparency is used (non-zero) for the text in Inkscape, but the package 'transparent.sty' is not loaded}%
    \renewcommand\transparent[1]{}%
  }%
  \providecommand\rotatebox[2]{#2}%
  \newcommand*\fsize{\dimexpr\f@size pt\relax}%
  \newcommand*\lineheight[1]{\fontsize{\fsize}{#1\fsize}\selectfont}%
  \ifx\svgwidth\undefined%
    \setlength{\unitlength}{118.80260558bp}%
    \ifx\svgscale\undefined%
      \relax%
    \else%
      \setlength{\unitlength}{\unitlength * \real{\svgscale}}%
    \fi%
  \else%
    \setlength{\unitlength}{\svgwidth}%
  \fi%
  \global\let\svgwidth\undefined%
  \global\let\svgscale\undefined%
  \makeatother%
  \begin{picture}(1,0.81789057)%
    \lineheight{1}%
    \setlength\tabcolsep{0pt}%
    \put(0,0){\includegraphics[width=\unitlength,page=1]{fig1.pdf}}%
    \put(0.15268208,0.78112254){\color[rgb]{0,0,0}\makebox(0,0)[lt]{\lineheight{0}\smash{\begin{tabular}[t]{l}rf antenna\end{tabular}}}}%
    \put(0,0){\includegraphics[width=\unitlength,page=2]{fig1.pdf}}%
    \put(0.95404584,0.13270557){\color[rgb]{0,0,0}\makebox(0,0)[t]{\lineheight{1.20000005}\smash{\begin{tabular}[t]{c}$g_\text{Ti}$\end{tabular}}}}%
    \put(-0.00082366,0.11376606){\color[rgb]{0,0,0}\makebox(0,0)[t]{\lineheight{1.20000005}\smash{\begin{tabular}[t]{c}$g_\text{Ca}$\end{tabular}}}}%
    \put(0.2832611,0.00644451){\color[rgb]{0,0.50196078,0}\makebox(0,0)[lt]{\lineheight{1.20000005}\smash{\begin{tabular}[t]{l}\textit{$B$}\end{tabular}}}}%
    \put(0,0){\includegraphics[width=\unitlength,page=3]{fig1.pdf}}%
    \put(0.59259887,0.14533031){\color[rgb]{0,0,0}\makebox(0,0)[lt]{\lineheight{1.20000005}\smash{\begin{tabular}[t]{l}\scriptsize{3/2}\end{tabular}}}}%
    \put(0.68729376,0.1958343){\color[rgb]{0,0,0}\makebox(0,0)[lt]{\lineheight{1.20000005}\smash{\begin{tabular}[t]{l}\scriptsize{5/2}\end{tabular}}}}%
    \put(0.78198829,0.21477357){\color[rgb]{0,0,0}\makebox(0,0)[lt]{\lineheight{1.20000005}\smash{\begin{tabular}[t]{l}\scriptsize{7/2}\end{tabular}}}}%
    \put(0.8703697,0.23371523){\color[rgb]{0,0,0}\makebox(0,0)[lt]{\lineheight{1.20000005}\smash{\begin{tabular}[t]{l}\scriptsize{9/2}\end{tabular}}}}%
    \put(0.71080603,0.33810747){\color[rgb]{0,0,0}\makebox(0,0)[lt]{\lineheight{1.20000005}\smash{\begin{tabular}[t]{l}\Large{$^4\text{F}_J$}\end{tabular}}}}%
    \put(0,0){\includegraphics[width=\unitlength,page=4]{fig1.pdf}}%
    \put(0.10535017,0.24028587){\color[rgb]{0,0,0}\makebox(0,0)[lt]{\lineheight{1.20000005}\smash{\begin{tabular}[t]{l}$\omega_\text{Ca}$\end{tabular}}}}%
    \put(0.52768826,0.23337383){\color[rgb]{0,0,0}\makebox(0,0)[lt]{\lineheight{1.20000005}\smash{\begin{tabular}[t]{l}$\omega_\text{Ti}$\end{tabular}}}}%
    \put(0,0){\includegraphics[width=\unitlength,page=5]{fig1.pdf}}%
    \put(0.42268755,0.51447568){\color[rgb]{0.83921569,0.15294118,0.15686275}\makebox(0,0)[lt]{\lineheight{1.20000005}\smash{\begin{tabular}[t]{l}$^{48}$Ti$^+$\end{tabular}}}}%
    \put(0.15023039,0.51447568){\color[rgb]{0.12156863,0.46666667,0.70588235}\makebox(0,0)[lt]{\lineheight{1.20000005}\smash{\begin{tabular}[t]{l}$^{40}$Ca$^+$\end{tabular}}}}%
    \put(0,0){\includegraphics[width=\unitlength,page=6]{fig1.pdf}}%
  \end{picture}%
\endgroup%

%% file: fig2.pdf_tex
\begingroup%
  \makeatletter%
  \providecommand\color[2][]{%
    \errmessage{(Inkscape) Color is used for the text in Inkscape, but the package 'color.sty' is not loaded}%
    \renewcommand\color[2][]{}%
  }%
  \providecommand\transparent[1]{%
    \errmessage{(Inkscape) Transparency is used (non-zero) for the text in Inkscape, but the package 'transparent.sty' is not loaded}%
    \renewcommand\transparent[1]{}%
  }%
  \providecommand\rotatebox[2]{#2}%
  \newcommand*\fsize{\dimexpr\f@size pt\relax}%
  \newcommand*\lineheight[1]{\fontsize{\fsize}{#1\fsize}\selectfont}%
  \ifx\svgwidth\undefined%
    \setlength{\unitlength}{340.21122249bp}%
    \ifx\svgscale\undefined%
      \relax%
    \else%
      \setlength{\unitlength}{\unitlength * \real{\svgscale}}%
    \fi%
  \else%
    \setlength{\unitlength}{\svgwidth}%
  \fi%
  \global\let\svgwidth\undefined%
  \global\let\svgscale\undefined%
  \makeatother%
  \begin{picture}(1,0.30960462)%
    \lineheight{1}%
    \setlength\tabcolsep{0pt}%
    \put(0,0){\includegraphics[width=\unitlength,page=1]{fig2.pdf}}%
    \put(0.03306762,0.24868217){\color[rgb]{0,0,0}\makebox(0,0)[lt]{\lineheight{1.20000005}\smash{\begin{tabular}[t]{l}Ca$^+$\end{tabular}}}}%
    \put(0.03306762,0.07232097){\color[rgb]{0,0,0}\makebox(0,0)[lt]{\lineheight{1.20000005}\smash{\begin{tabular}[t]{l}Ti$^+$\end{tabular}}}}%
    \put(-0.00155012,0.1522232){\color[rgb]{0,0,0}\makebox(0,0)[lt]{\lineheight{1.20000005}\smash{\begin{tabular}[t]{l}motion\end{tabular}}}}%
    \put(0.84873739,0.28395415){\color[rgb]{0,0,0}\makebox(0,0)[lt]{\lineheight{1.20000005}\smash{\begin{tabular}[t]{l}RAP\end{tabular}}}}%
    \put(0.92413603,0.29277234){\color[rgb]{0,0,0}\makebox(0,0)[lt]{\lineheight{1.20000005}\smash{\begin{tabular}[t]{l}FD\end{tabular}}}}%
    \put(0.59521839,0.29277234){\color[rgb]{0,0,0}\makebox(0,0)[lt]{\lineheight{1.20000005}\smash{\begin{tabular}[t]{l}FD\end{tabular}}}}%
    \put(0.34169933,0.28395415){\color[rgb]{0,0,0}\makebox(0,0)[lt]{\lineheight{1.20000005}\smash{\begin{tabular}[t]{l}rf\end{tabular}}}}%
    \put(0.38358496,0.10759296){\color[rgb]{0,0,0}\makebox(0,0)[lt]{\lineheight{1.20000005}\smash{\begin{tabular}[t]{l}rf\end{tabular}}}}%
    \put(0.50608313,0.10759296){\color[rgb]{0,0,0}\makebox(0,0)[lt]{\lineheight{1.20000005}\smash{\begin{tabular}[t]{l}rf\end{tabular}}}}%
    \put(0.54892355,0.28395415){\color[rgb]{0,0,0}\makebox(0,0)[lt]{\lineheight{1.20000005}\smash{\begin{tabular}[t]{l}rf\end{tabular}}}}%
    \put(0,0){\includegraphics[width=\unitlength,page=2]{fig2.pdf}}%
    \put(0.24249638,0.1693195){\color[rgb]{0,0,0}\makebox(0,0)[lt]{\lineheight{1.20000005}\smash{\begin{tabular}[t]{l}state\\prep.\end{tabular}}}}%
    \put(0.1322707,0.19356926){\color[rgb]{0,0,0}\makebox(0,0)[lt]{\lineheight{1.20000005}\smash{\begin{tabular}[t]{l}GSC\end{tabular}}}}%
    \put(0.68339893,0.19356926){\color[rgb]{0,0,0}\makebox(0,0)[lt]{\lineheight{1.20000005}\smash{\begin{tabular}[t]{l}GSC\end{tabular}}}}%
    \put(0.77025317,0.10799522){\color[rgb]{0,0,0}\makebox(0,0)[lt]{\lineheight{1.20000005}\smash{\begin{tabular}[t]{l}RSB\end{tabular}}}}%
    \put(0,0){\includegraphics[width=\unitlength,page=3]{fig2.pdf}}%
    \put(0.41885732,0.00618584){\color[rgb]{0,0,0}\makebox(0,0)[lt]{\lineheight{1.20000005}\smash{\begin{tabular}[t]{l}\SI{300}{\micro\second}\end{tabular}}}}%
    \put(0,0){\includegraphics[width=\unitlength,page=4]{fig2.pdf}}%
  \end{picture}%
\endgroup%